\documentclass[aps,11pt,twoside,tightenlines,superscriptaddress]{revtex4}
\usepackage{amsmath,amssymb,amsthm,amsfonts,graphicx}
\usepackage[colorlinks=false]{hyperref}

\newcommand{\ket}[1]{| #1 \rangle}        
\newcommand{\bra}[1]{\langle #1 |}        
\newcommand{\scalar}[2]{\left\langle #1 | #2 \right\rangle}        

\newcommand{\ii}{\mathbb{I}}
\newcommand{\norm}[1]{\left\| #1\right\|}        
\newcommand{\deriv}[2]{\frac{\textrm{d}#1}{\textrm{d}#2}}        
\newcommand{\bigabs}[1]{\Big|\!#1\!\Big|}        
\newcommand{\bignorm}[1]{\Big\|#1\Big\|}        

\newtheorem{theorem}{Theorem}
\newtheorem{lemma}{Lemma}

\begin{document}


\title{How to Make the Quantum Adiabatic Algorithm Fail}
\author{Edward Farhi}
\email{farhi@mit.edu}
\affiliation{Center for Theoretical Physics, Massachusetts Institute of Technology, 
Cambridge, Massachusetts, 02139}
\author{Jeffrey Goldstone}
\affiliation{Center for Theoretical Physics, Massachusetts Institute of Technology, 
Cambridge, Massachusetts, 02139}
\author{Sam Gutmann}
\affiliation{Department of Mathematics, Northeastern University, Boston, Massachusetts, 02115}
\author{Daniel Nagaj}
\affiliation{Center for Theoretical Physics, Massachusetts Institute of Technology, 
Cambridge, Massachusetts, 02139}

\date{Dec 19, 2005}

\begin{abstract}
The quantum adiabatic algorithm is a Hamiltonian based quantum algorithm designed to find
the minimum of a classical cost function whose domain has size $N$. We show that
poor choices for the Hamiltonian can guarantee that the algorithm will not find the minimum
if the run time grows more slowly than $\sqrt{N}$. These poor choices are nonlocal
and wash out any structure in the cost function to be minimized and the best 
that can be hoped for is Grover speedup. These failures tell us what not to do when 
designing quantum adiabatic algorithms.
\end{abstract}

\maketitle


\section{Introduction}

The quantum adiabatic algorithm was introduced \cite{adiabatic} as a quantum algorithm for 
finding the minimum
of a classical cost function $h(z)$, where $z=0,\dots,N-1$. This cost function is used to define
a quantum Hamiltonian diagonal in the $z$ basis:
\begin{eqnarray}
	H_P = \sum_{z=0}^{N-1} h(z) \ket{z}. \label{problemHam}
\end{eqnarray}
The goal is now to find the ground state of $H_P$. To this end a ``beginning'' Hamiltonian $H_B$ is 
introduced with a known and easy to construct ground state $\ket{g_B}$. The quantum computer
is a system governed by the time dependent Hamiltonian 
\begin{eqnarray}
	H(t) = (1-t/T) H_B + (t/T) H_P, \label{adiabaticHam}
\end{eqnarray}
where $T$ controls the rate of change of $H(t)$. Note that $H(0)=H_B$ and $H(T)=H_P$. 
The state of the system obeys the Schr\"odinger equation,
\begin{eqnarray}
	i \deriv{}{t} \ket{\psi(t)} = H(t) \ket{\psi(t)}, \label{Schrodinger}
\end{eqnarray}
where we choose
\[
	\ket{\psi(0)} = \ket{g_B}
\]
and run the algorithm for time $T$. By the adiabatic theorem, if $T$ is large enough then $\ket{\psi(T)}$
will have a large component in the ground state subspace of $H_P$. (Note we are not bothering to state the 
necessary condition on the lack of degeneracy of the spectrum of $H(t)$ for $0<t<T$, since it will
not play a role in the results we establish in this paper.) A measurement of $z$ 
can then be used to find the minimum of $h(z)$. The algorithm is useful if the required
run time $T$ is not too large as a function of $N$.

There is hope that there may be combinatorial search problems, defined on $n$ bits so that $N=2^n$, where 
for certain ``interesting'' subsets of the instances the run time $T$ grows subexponentially in $n$. 
A positive result of this kind would greatly expand the known power of quantum computers. At the same time
it is worthwhile to understand the circumstances under which the algorithm is doomed to fail.

In this paper we prove some general results which show that with certain choices of $H_B$ or $H_P$ the
algorithm will not succeed if $T$ is $o(\sqrt{N})$, that is $T/\sqrt{N}\rightarrow 0$ as $N\rightarrow\infty$,
so that improvement beyond Grover speedup is impossible. 
We view these failures as due to poor choices for $H_B$ and $H_P$, which teach us what not to do
when looking for good algorithms. 
We guarantee failure by removing any structure which might exist in $h(z)$ from either $H_B$ or $H_P$. 
By structure we mean that $z$ is written as a bit string and both $H_B$ and $H_P$ are sums of 
terms involving only a few of the corresponding qubits.

In Section II we show that regardless of the form of $h(z)$ if $H_B$ is a one dimensional projector
onto the uniform superposition of all the basis states $\ket{z}$, then the quantum adiabatic algorithm fails.
Here all the $\ket{z}$ states are treated identically by $H_B$ so any structure contained in $h(z)$ is
lost in $H_B$. In Section III we consider a scrambled $H_P$ that we get by replacing the cost function
$h(z)$ by $h(\pi(z))$ where $\pi$ is a permutation of $0$ to $N-1$. Here the values of $h(z)$ and 
$h(\pi(z))$ are the same but the relationship between input and output is scrambled by the permutation.
This effectively destroys any structure in $h(z)$ and typically results in algorithmic failure. 

The quantum adiabatic algorithm is a special case of Hamiltonian based continuous time quantum algorithms, 
where the quantum state obeys (\ref{Schrodinger}) and the algorithm consists of specifying $H(t)$, the initial
state $\ket{\psi(0)}$, a run time $T$ and the operators to be measured at the end of the run.
In the Hamiltonian language, the Grover problem can be recast as the problem of finding the ground state
of
\begin{eqnarray}
	H_w = E(\ii-\ket{w}\bra{w}), \label{groverHam}
\end{eqnarray}
where $w$ lies between $0$ and $N-1$. The algorithm designer can apply $H_w$, but in this oracular setting,
$w$ is not known. In reference \cite{analog} the following result was proved. Let 
\begin{eqnarray}
	H(t) = H_D(t) + H_w, \label{analogHam}
\end{eqnarray}
where $H_D$ is any time dependent ``driver'' Hamiltonian independent of $w$. Assume also that the initial
state $\ket{\psi(0)}$ is independent of $w$. For each $w$ we want the algorithm to be successful,
that is $\ket{\psi(T)}=\ket{w}$. It then follows that
\begin{eqnarray}
	T \geq \frac{\sqrt{N}}{2E}. \label{Groverscaling}
\end{eqnarray}
The proof of this result is a continuous-time version of the BBBV oracular proof \cite{BBBV}.
Our proof techniques in this paper are similar to the methods used to prove the result just stated.


\section{General search starting with a one-dimensional projector}

In this section we consider a completely general cost function $h(z)$ with
$z=0,\dots,N-1$. The goal is to use the quantum adiabatic algorithm to find the ground
state of $H_P$ given by (\ref{problemHam}) with $H(t)$ given by (\ref{adiabaticHam}). Let
\begin{eqnarray}
	\ket{s} = \frac{1}{\sqrt{N}} \sum_{z=0}^{N-1} \ket{z}
\end{eqnarray}
be the uniform superposition over all possible values $z$. If we pick
\begin{eqnarray}
	H_B = E (\ii- \ket{s}\bra{s}) \label{groverstart}
\end{eqnarray}
and $\ket{\psi(0)}=\ket{s}$, then the adiabatic algorithm fails in the following sense: 

\begin{theorem}
Let $H_P$ be diagonal in the $z$ basis with a ground state subspace of dimension $k$. Let
\[
	H(t) = (1-t/T) E \left( \ii - \ket{s}\bra{s}\right) + (t/T) H_P.
\]
Let $P$ be the projector onto the ground state subspace of $H_P$ and let $b>0$ be the success
probability, that is, $b=\bra{\psi(T)}P\ket{\psi(T)}$. Then
\[
	T \geq \frac{b}{E}\sqrt{\frac{N}{k}} - \frac{2\sqrt{b}}{E}.
\]
\end{theorem}

\begin{proof}
Keeping $H_P$ fixed, we introduce $N-1$ additional beginning Hamiltonians as follows.
For $x=0,\dots,N-1$ let $V_x$ be a unitary operator diagonal in the $z$ basis with
\[
	\bra{z}V_x\ket{z} = e^{2\pi i z x /N}
\]
and let
\[
	\ket{x} = V_x \ket{s} = \frac{1}{\sqrt{N}} \sum_{z=0}^{N-1} e^{2\pi i z x /N}\ket{z}
\]
so that the $\{\ket{x}\}$ form an orthonormal basis. Note also that
\[
	\ket{x=0}=\ket{s}.
\]
We now define 
\[
  H_x(t) = ( 1-t/T ) E (\ii-\ket{x}\bra{x}) + (t/T) H_P,
\]
with corresponding evolution operator $U_x (t_2,t_1)$. 
Note that $H(t)$ above is $H_0(t)$ with the corresponding evolution operator $U_0$. For each $x$ we 
evolve with $H_x(t)$ from the ground state of $H_x(0)$, which is $\ket{x}$. Note that 
$H_x=V_x H_0 V_x^{\dagger}$ and $U_x=V_x U_0 V_x^{\dagger}$. Let $\ket{f_x}=U_x(T,0)\ket{x}$. For
each $x$ the success probability is $\bra{f_x}P\ket{f_x}$, which is equal to $b$ since $P$ commutes
with $V_x$.
The key point is that if we run the Hamiltonian evolution with $H_x$ backwards in time, 
we would then be finding $x$, that is, solving the Grover problem.
However, this should not be possible unless the run time $T$ is of order $\sqrt{N}$. 

Let $U_R$ be the evolution operator corresponding to an $x$-independent reference Hamiltonian
\[
  H_R(t) = (1-t/T) E + (t/T) H_P.
\]
Let $\ket{g_x} = \frac{1}{\sqrt{b}}P\ket{f_x}$ be the normalized component of $\ket{f_x}$ in the
ground state subspace of $H_P$. We consider the difference in backward evolution from $\ket{g_x}$
with Hamiltonians $H_x$ and $H_R$, and sum on $x$,
\[
  S(t) = \sum_x \norm{U_x^{\dagger}(T,t)\ket{g_x}-U_R^{\dagger}(T,t)\ket{g_x}}^2.
\]
Clearly $S(T)=0$, and
\begin{eqnarray*}
  S(0) &=& \sum_x \norm{ U_x^{\dagger}(T,0)\ket{g_x} - U_R^{\dagger}(T,0)\ket{g_x} }^2.
\end{eqnarray*}
Now $\ket{g_x}=\sqrt{b}\ket{f_x}+\sqrt{1-b}\ket{f_x^{\perp}}$ where $\ket{f_x^{\perp}}$ is orthogonal
to $\ket{f_x}$. Since $U_x^{\dagger}(T,0)\ket{f_x}=\ket{x}$ we have
\begin{eqnarray*}
       S(0) = \sum_x \norm{ \sqrt{b}\ket{x} + \sqrt{1-b}\ket{x^\perp} - \ket{i_x} }^2,
\end{eqnarray*}
where for each $x$, $\ket{x^\perp}$ and $\ket{i_x}$ are normalized states with $\ket{x^\perp}$ orthogonal 
to $\ket{x}$. Since $H_R$ commutes with $H_P$, $\ket{i_x}=U_R^{\dagger}(T,0)\ket{g_x}$ 
is an element of the $k$-dimensional ground state subspace of $H_P$. We have
\begin{eqnarray*}
  S(0) &=& 2N - \sum_x \left[ \sqrt{b} \scalar{x}{i_x} + \sqrt{1-b}\langle x^\perp | i_x\rangle+c.c.\right] \\
	&\geq& 2N - 2\sqrt{b} \sum_x \bigabs{\scalar{x}{i_x}} - 2N\sqrt{1-b}.
\end{eqnarray*}
Choosing a basis $\{\ket{G_j}\}$ for the $k$ dimensional ground state subspace of $H_P$ and writing 
$\ket{i_x} = a_{x1}\ket{G_1} + \cdots + a_{xk}\ket{G_k}$ gives
\begin{eqnarray}
  \sum_x \bigabs{\scalar{x}{i_x}} 
  &\leq& \sum_{x,j} \left|a_{xj}\right| \cdot \bigabs{\scalar{x}{G_j}} \label{boundix} \\
  &\leq& \sqrt{ \sum_{x,j} \left|a_{xj}\right|^2 \sum_{x',j'} \bigabs{\scalar{x'}{G_{j'}}}^2 } 
	= \sqrt{Nk}. \nonumber
\end{eqnarray}
Thus
\begin{eqnarray}
  S(0)\geq 2N(1-\sqrt{1-b})- 2\sqrt{b}\sqrt{Nk}. \label{s0eqn}
\end{eqnarray}
We will use the Schr\"{o}dinger equation to find the time derivative of $S(t)$:
\begin{eqnarray*}
  \deriv{}{t}S(t) &=& -\sum_x \deriv{}{t} \left[ \bra{g_x}U_x(T,t)U_R^{\dagger}(T,t) \ket{g_x} + c.c. \right]  \\
	&=& -i \sum_x \bra{g_x}U_x(T,t)[H_x(t)-H_R(t)]U_R^{\dagger}(T,t) \ket{g_x} + c.c. \\
	&=& -2\,\textrm{Im} \sum_x (1-t/T)E \bra{g_x}U_x(T,t) \ket{x}\bra{x} U_R^{\dagger}(T,t) \ket{g_x}.
\end{eqnarray*}
Now 
\begin{eqnarray*}
	\left| \deriv{}{t} S(t) \right| &\leq& 2E (1-t/T) \sum_x
		\left| \bra{g_x}U_x(T,t) \ket{x}\bra{x} U_R^{\dagger}(T,t) \ket{g_x} \right| \\
	&\leq& 2E (1-t/T) \sum_x
		\left| \bra{x} U_R^{\dagger}(T,t) \ket{g_x} \right|.
\end{eqnarray*}
Using the same technique as in (\ref{boundix}), we obtain
\begin{eqnarray*}
	\left| \deriv{}{t} S(t) \right| &\leq& 2E (1-t/T) \sqrt{Nk}.
\end{eqnarray*}
Therefore
\begin{eqnarray*}
	\int^T_0 \left| \deriv{}{t} S(t) \right| \textrm{d}t \leq ET \sqrt{Nk}.
\end{eqnarray*}
Now $S(0)\leq S(T) + \int^T_0 \left| \deriv{}{t} S(t) \right| \textrm{d}t$ and $S(T)=0$ so 
\[
	S(0)\leq ET\sqrt{Nk}.
\]
Combining this with (\ref{s0eqn}) gives
\[
   ET\sqrt{Nk} \geq 2N (1-\sqrt{1-b}) - 2\sqrt{b} \sqrt{Nk},
\]
which implies what we wanted to prove:
\[
   T \geq \frac{b}{E}\sqrt{\frac{N}{k}} - \frac{2\sqrt{b}}{E}.
\]
\end{proof}

How do we interpret Theorem 1? The goal is to find the minimum of the cost function $h(z)$
using the quantum adiabatic algorithm. It is natural to pick for $H_B$ a Hamiltonian
whose ground state is $\ket{s}$, the uniform superposition of all $\ket{z}$ states. However
if we pick $H_B$ to be the one dimensional projector $E(\ii-\ket{s}\bra{s})$ the algorithm
will not find the ground state if $T/\sqrt{N}$ goes to $0$ as $N$ goes to infinity. 
The problem is that $H_B$ has no structure and makes no reference to $h(z)$. 
Our hope is that the algorithm might be useful for interesting computational problems 
if $H_B$ has structure that reflects the form of $h(z)$. 

Note that Theorem 1 explains the algorithmic failure discovered by 
\v{Z}nidari\v{c} and Horvat \cite{znidaric} for a particular set of $h(z)$.

For a simple but convincing example of the importance of the choice of $H_B$, suppose we take
a decoupled $n$ bit problem which consists of $n$ clauses each acting on one bit, say for each bit $j$ 
\begin{eqnarray*}
  h_j(z)=\left\{
    \begin{array}{rl}
	0 & \quad \textrm{if} \, z_{j}=0, \\
	1 & \quad \textrm{if} \, z_{j}=1,
    \end{array}
  \right. 
\end{eqnarray*}
so 
\begin{eqnarray}
	h(z)=z_1+z_2+\dots+z_n  \label{decoupledcost}.
\end{eqnarray}
Let us pick a beginning Hamiltonian reflecting the bit structure of the problem,
\begin{eqnarray}
	H_B = \sum_{j=1}^{n} \frac{1}{2}\left(1-\sigma_x^{(j)}\right). \label{sumSX}
\end{eqnarray}
The ground state of $H_B$ is $\ket{s}$,
The quantum adiabatic algorithm acts on each bit independently, producing a success probability of
\[
	p = \left(1-q(T)\right)^n,
\]
where $q(T)\rightarrow 0$ as $T\rightarrow\infty$ is the transition probability between the ground state
and the excited state of a single qubit. As long as $n q(T)\rightarrow const.$ we have a constant
probability of success. This can be achieved for $T$ of order $\sqrt{n}$, because for a two level system
with a nonzero gap, the probability of a transition is $q(T)=O(T^{-2})$. (For details, see Appendix A.)
However, from Theorem 1 we see that a poor choice of $H_B$ would make the quantum adiabatic algorithm 
fail on this simple decoupled $n$ bit problem by destroying the bit structure. 

Next, suppose the satisfiability problem we are trying to solve has clauses involving 
say 3 bits. If clause $c$ involves bits $i_c$, $j_c$ and $k_c$ we may define the clause cost 
function
\begin{eqnarray*}
  h_c(z)=\left\{
    \begin{array}{rl}
	0 & \quad \textrm{if} \,\, z_{i_c},z_{j_c},z_{k_c} \, \textrm{satisfy clause} \, c, \\
	1 & \quad \textrm{otherwise}.
    \end{array}
  \right.
\end{eqnarray*}
The total cost function is then
\begin{eqnarray*}
  h(z) = \sum_c h_c(z).
\end{eqnarray*}
To get $H_B$ to reflect the bit and clause structure we may pick
\begin{eqnarray*}
  H_{B,c} = \frac{1}{2}\left[(1-\sigma_x^{(i_c)}) + (1-\sigma_x^{(j_c)}) + (1-\sigma_x^{(k_c)}) \right]
\end{eqnarray*}
with
\begin{eqnarray}
  H_B = \sum_c H_{B,c}. \label{clauseham}
\end{eqnarray}
In this case the ground state of $H_B$ is again $\ket{s}$. With this setup, Theorem 1 does not apply.

\begin{figure}
	\begin{center}
	\includegraphics[width=4.5in]{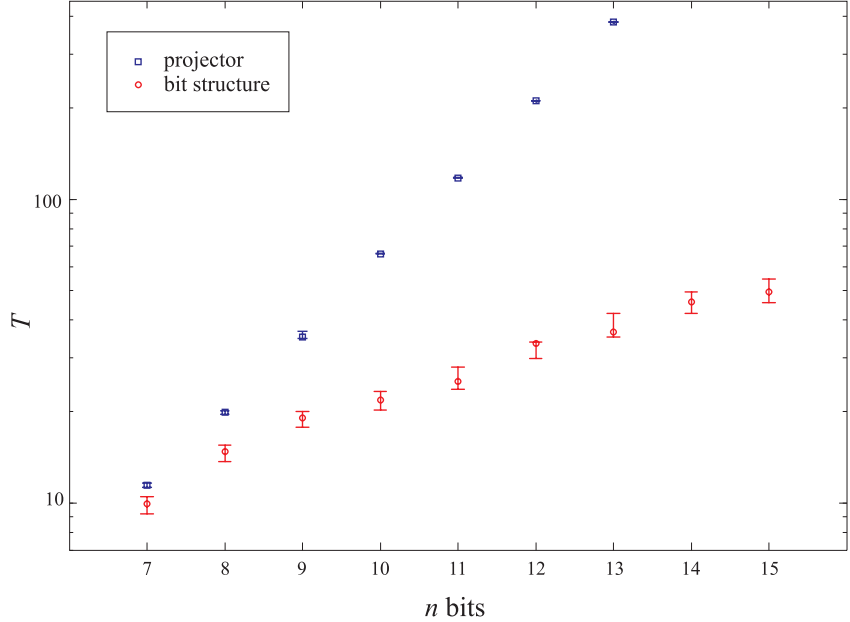}
	\end{center}
	\caption{Median required run time $T$ versus bit number. At each bit number there are 50 random instances
 		 of Exact Cover with a single satisfying assignment. We choose the required run time to be 
		 the value of $T$ for which quantum adiabatic algorithm 
 		 has success probability between 0.2 and 0.21.
		 For the projector beginning Hamiltonian we use \eqref{groverstart} with $E=n/2$.
	 	 The plot is log-linear. The error bars show the 95\% confidence interval for the true medians.}
	\label{loglin}
\end{figure}

We did a numerical study of a particular satisfiability problem, Exact Cover. For this problem 
if clause $c$ involves bits $i_c$, $j_c$ and $k_c$, the cost function is 
\begin{eqnarray*}
  h_c(z)=\left\{
    \begin{array}{rl}
	0 & \quad \textrm{if} \,\, z_{i_c}+z_{j_c}+z_{k_c}=1, \\
	1 & \quad \textrm{otherwise}.
    \end{array}
  \right.
\end{eqnarray*}
Some data is presented in FIG. 1. Here we see that with a structured beginning Hamiltonian 
the required run times are substantially lower than with the projector $H_B$.


\section{Search with a scrambled problem hamiltonian}

In the previous section we showed that removing all structure from $H_B$ dooms the 
quantum adiabatic algorthm to failure. In this section we remove structure from the problem to be
solved ($H_P$) and show that this leads to algorithmic failure. Let $h(z)$ be a cost function 
whose minumum we seek. Let $\pi$ be a permutation of $0,1,\dots,N-1$ and let
\[
	h^{[\pi]}(z)=h\left(\pi^{-1}(z)\right).
\]
We will show that no continuous time quantum algorithm (of a very general form) can find the minimum
of $h^{[\pi]}$ for even a small fraction of all $\pi$ if $T$ is $o(\sqrt{N})$. Classically, this problem
takes order $N$ calls to an oracle.

Without loss of generality let $h(0)=0$, and $h(1),h(2),\dots,h(N-1)$ all be positive. 
For any permutation $\pi$ of $0,1,\dots,N-1$ we define a problem Hamiltonian $H_{P,\pi}$, 
diagonal in the $z$ basis, as 
\[
	H_{P,\pi} = \sum_{z=0}^{N-1} h^{[\pi]}(z) \ket{z}\bra{z} 
		=\sum_{z=0}^{N-1} h(z) \ket{\pi(z)}\bra{\pi(z)}.
\]
Now consider the Hamiltonian 
\begin{eqnarray}
	H_{\pi}(t)=H_D(t)+c(t) H_{P,\pi} \label{continuoustimealgorithm}
\end{eqnarray}
for an arbitrary $\pi$-independent driving Hamiltonian
$H_D(t)$ with $|c(t)|\leq1$ for all $t$. Using this composite Hamiltonian, 
we evolve the $\pi$-independent starting state $\ket{\psi(0)}$ for time $T$, reaching the state 
$\ket{\psi_\pi(T)}$. This setup is more general than the quantum adiabatic algorithm since we do not
require $H_D(t)$ or $c(t)$ to be slowly varying. 
Success is achieved if the overlap of $\ket{\psi_{\pi}(T)}$ 
with $\ket{\pi(0)}$ is large.

We first show
\begin{lemma}
\begin{eqnarray}
	\sum_{\pi,\pi'} \Big\| \ket{\psi_\pi(T)}-\ket{\psi_{\pi'}(T)} \Big\|^2 \leq 4h^* T N!\sqrt{N-1}, \label{permutationlemma}
\end{eqnarray}
where the sum is over all pairs of permutations $\pi,\pi'$ that differ by a single transposition 
involving $\pi(0)$, and $h^*=\sqrt{\sum h(z)^2 / (N-1)}$. 
\end{lemma}

\begin{proof}
For two different permutations $\pi$ and $\pi'$ let $\ket{\psi_{\pi}(t)}$ be the state obtained
by evolving from $\ket{\psi(0)}$ with $H_{\pi}$ and let $\ket{\psi_{\pi'}(t)}$ be the state obtained
by evolving from $\ket{\psi(0)}$ with $H_{\pi'}$.

Now
\begin{eqnarray*}
  \deriv{}{t} \Big\| \ket{\psi_\pi(t)}-\ket{\psi_{\pi'}(t)} \Big\|^2 
	&=& -\deriv{}{t} \scalar{\psi_{\pi}(t)}{\psi_{\pi'}(t)} + c.c. \\
	&=& i \bra{\psi_{\pi}(t)}(H_{\pi}(t)-H_{\pi'}(t))\ket{\psi_{\pi'}(t)} + c.c. \\
	&\leq& 2 \Big| \bra{\psi_{\pi}(t)}(H_{\pi}(t)-H_{\pi'}(t))\ket{\psi_{\pi'}(t)} \Big|.
\end{eqnarray*}
We now consider the case when $\pi$ and $\pi'$ differ by a single transposition involving $\pi(0)$. 
Specifically, $\pi'=\pi \circ (a\leftrightarrow 0)$ for some $a$. 
Now if $\pi(0) = i$ and $\pi(a) =j$, we have $\pi'(0) =j$ and $\pi'(a) =i$. 
Therefore, since $h(0)=0$, 
\begin{eqnarray*}
  H_{P,\pi}-H_{P,\pi'} = c(t) h(a) \left( \ket{j}\bra{j} - \ket{i}\bra{i} \right)
			= c(t) h(a) \left( \ket{\pi(a)}\bra{\pi(a)} - \ket{\pi'(a)}\bra{\pi'(a)} \right),
\end{eqnarray*}
so we can write
\begin{eqnarray*}
  \deriv{}{t} \sum_{\pi,\pi'} \Big\| \ket{\psi_\pi(t)}-\ket{\psi_{\pi'}(t)} \Big\|^2 
  &\leq&
	2 |c(t)| \sum_{\pi,\pi'} h(a) \Big| \bra{\psi_{\pi}(t)} 
		\left( \ket{\pi(a)}\bra{\pi(a)} - \ket{\pi'(a)}\bra{\pi'(a)} \right)  
		\ket{\psi_{\pi'}(t)} \Big|.
\end{eqnarray*}
This further simplifies to
\begin{eqnarray*}
  \deriv{}{t} \sum_{\pi,\pi'} \Big\| \ket{\psi_\pi(t)}-\ket{\psi_{\pi'}(t)} \Big\|^2
  &\leq&
	2 \sum_{\pi,\pi'} h(a) \left(
		\bigabs{\scalar{\psi_{\pi}(t)}{\pi(a)}} +
		\bigabs{\scalar{\pi'(a)}{\psi_{\pi'}(t)}} \right) \\
  &=&
	2 \sum_{\pi} \sum_{a\neq 0} h(a) 
		\bigabs{\scalar{\psi_{\pi}(t)}{\pi(a)}} +
	2 \sum_{\pi'} \sum_{a\neq 0} h(a) 
		\bigabs{\scalar{\pi'(a)}{\psi_{\pi'}(t)}}  \\
  &=&
	4 \sum_{\pi} \sum_{a\neq 0} h(a) \bigabs{\scalar{\psi_{\pi}(t)}{\pi(a)}} \\
  &=&
	4 \sum_{\pi} \sum_{a} h(a) \bigabs{\scalar{\psi_{\pi}(t)}{\pi(a)}} \\
  &\leq&
	4 \sum_{\pi} \sqrt{\sum_a h(a)^2} = 4h^* N! \sqrt{N-1} .
\end{eqnarray*}
where we used the Cauchy-Schwartz inequality to obtain the last line. Integrating this inequality
for time $T$, we obtain the result we wanted to prove,
\begin{eqnarray*}
  \sum_{\pi,\pi'} \Big\| \ket{\psi_\pi(T)}-\ket{\psi_{\pi'}(T)} \Big\|^2 &\leq& 4 h^* T N! \sqrt{N-1},
\end{eqnarray*}
where the sum is over $\pi$ and $\pi'$ differing by a single transposition involving $\pi(0)$.
\end{proof}

Next we establish
\begin{lemma}
  Suppose $\ket{1}$, $\ket{2}$, $\ket{L}$ are orthonormal vectors and $\bigabs{\scalar{\psi_i}{i}}^2 \geq b$ for 
normalized vectors $\ket{\psi_i}$, where $i=1,\dots,L$. Then for any normalized $\ket{\varphi}$,
\begin{eqnarray}
   \sum_{i=1}^{L} \bignorm{\ket{\psi_i}-\ket{\varphi}}^2 \geq bL-2\sqrt{L}. \label{vectorlemma}
\end{eqnarray}
\end{lemma}
\begin{proof}
Write
\begin{eqnarray*}
  \sum_i \bignorm{\ket{\psi_i}-\ket{\varphi}}^2 &\geq& \sum_i \bigabs{\scalar{i}{\psi_i}-\scalar{i}{\varphi}}^2 \\
	&\geq& \sum_i \bigabs{\scalar{i}{\psi_i}}^2 - 2 \sum_i \bigabs{\scalar{i}{\psi_i}} \bigabs{\scalar{i}{\varphi}}
\end{eqnarray*}
and use the Cauchy-Schwartz inequality to obtain
\begin{eqnarray*}
  \sum_i \bignorm{\ket{\psi_i}-\ket{\varphi}}^2 &\geq& b L 
	- 2 \sqrt{\sum_i \bigabs{\scalar{i}{\psi_i}}^2 } \sqrt{\sum_i \bigabs{\scalar{i}{\varphi}}^2} \\
    &\geq& bL - 2 \sqrt{L}.
\end{eqnarray*}
\end{proof}

We are now ready to state the main result of this section. 
\begin{theorem}
  Suppose that a continuous time algorithm of the form \eqref{continuoustimealgorithm} 
  succeeds with probability at least $b$, i.e. 
$\bigabs{\scalar{\psi_{\pi}(T)}{\pi(0)}}^2\geq b$, for a set of $\epsilon N!$ permutations.
Then
\begin{eqnarray}
   T\geq \frac{\epsilon^2 b}{16 h^*} \sqrt{N-1} - \frac{\epsilon\sqrt{\epsilon/2}}{4h^*}.
\end{eqnarray}
\end{theorem}
\begin{proof}

For any permutation $\pi$, there are $N-1$ permutations $\pi'_a$ obtained from $\pi$
by first transposing $0$ and $a$. For each $\pi$ let $\mathcal{S}_{\pi}$ be the subset of those
$N-1$ permutations on which the algorithm succeeds with probability at least $b$. 
Any such permutation appears in exactly $N-1$ of the sets $\mathcal{S}_{\pi}$ so we have
\[
	\sum_\pi \left|\mathcal{S}_\pi\right| = (N-1)\epsilon N!.
\]
Let $M$ be the number of sets $\mathcal{S}_{\pi}$ with $\left|\mathcal{S}_{\pi}\right|\geq\frac{\epsilon}{2}(N-1)$. Now
\begin{eqnarray*}
  \sum_{\pi} \left| \mathcal{S}_{\pi} \right| &=& \sum_{\left| \mathcal{S}_{\pi} \right|\geq \frac{\epsilon}{2}(N-1)} \left| \mathcal{S}_{\pi} \right|
	+ \sum_{\left| \mathcal{S}_{\pi} \right|< \frac{\epsilon}{2}(N-1)} \left| \mathcal{S}_{\pi} \right| \\
    \sum_{\pi} \left| \mathcal{S}_{\pi} \right| &\leq& M(N-1) + (N!-M)\frac{\epsilon}{2}(N-1), \\
  (N-1)\epsilon N! &\leq& M(N-1) + N!\frac{\epsilon}{2}(N-1),
\end{eqnarray*}
so $M\geq\frac{\epsilon}{2}N!$, i.e. at least 
$\frac{\epsilon}{2} N!$ of the sets $\mathcal{S}_{\pi}$ must contain 
at least $\frac{\epsilon}{2}(N-1)$ permutations on which the algorithm succeeds with probability at least $b$.
For the corresponding $\pi$, we have
\[
   \sum_{\pi'_a} \bignorm{\ket{\psi_\pi (T)}-\ket{\psi_{\pi'_a} (T)}}^2  
	\geq b \frac{\epsilon}{2}(N-1) - 2\sqrt{\frac{\epsilon}{2}(N-1)}.
\]
by Lemma 2. (Note that the algorithm is not assumed to succeed with probability $b$ on $\pi$.) 
Since there are at least $\frac{\epsilon}{2}N!$ such $\pi$, 
\begin{eqnarray*}
  \sum_{\pi,\pi'} \bignorm{\ket{\psi_\pi (T)}-\ket{\psi_{\pi'}(T)}}^2 
	\geq \frac{\epsilon}{2}N! \left(b \frac{\epsilon}{2}(N-1) - 2\sqrt{\frac{\epsilon}{2}(N-1)}\right),
\end{eqnarray*}
where the sum is over all permutations $\pi$ and $\pi'$ which differ by a single transposition involving $\pi(0)$.
Combining this with Lemma 1 we obtain
\begin{eqnarray*}
   T\geq \frac{\epsilon^2 b}{16 h^*} \sqrt{N-1} - \frac{\epsilon\sqrt{\epsilon/2}}{4 h^*},
\end{eqnarray*}
which is what we wanted to prove.
\end{proof}

What we have just shown is that no continuous time algorithm of the form 
\eqref{continuoustimealgorithm} can find the minimum of $H_{P,\pi}$ with a 
constant success probability for even a fraction $\epsilon N!$ of all permutations $\pi$ 
if $T$ is $o(\sqrt{N})$. 
A typical permutation $\pi$ yields an $H_{P,\pi}$ with no structure relevant to 
any fixed $H_D$ and the algorithm can not find the ground state of $H_{P,\pi}$ efficiently. 

\begin{figure}
	\begin{center}
	\includegraphics[width=4.5in]{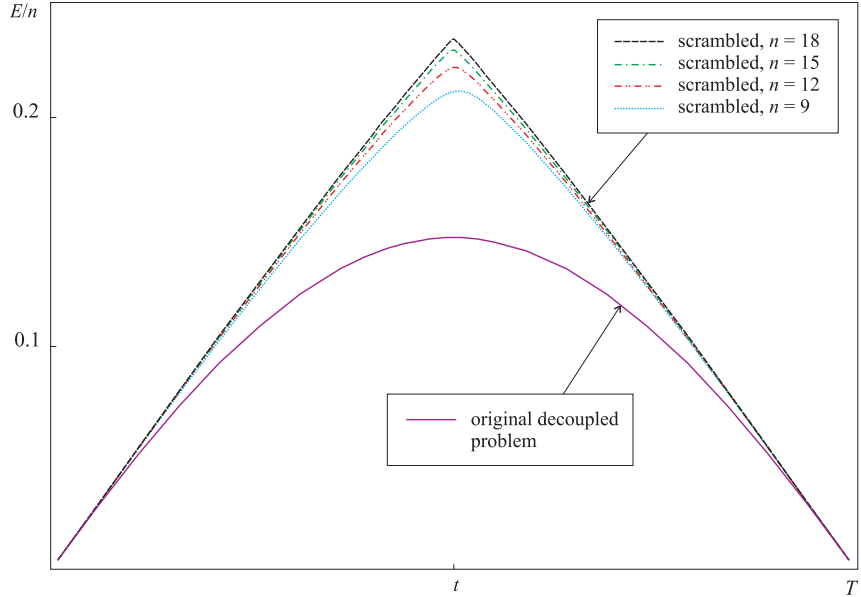}
	\end{center}
	\caption{The scaled ground state energy $E/n$ for a quantum adiabatic algorithm Hamiltonian
		 of a decoupled problem. The lowest curve corresponds to the original decoupled problem. 
		 The upper ``triangular'' curves correspond to single instances of 
		 the $n$-bit decoupled problem, where the problem Hamiltonian was scrambled.}
	\label{triangle}
\end{figure}

To illustrate the nature of this failure for the quantum adiabatic algorithm 
for a typical permutation, consider again the decoupled
$n$ bit problem with $h(z)$ given by \eqref{decoupledcost} and $H_B$ given by \eqref{sumSX}.
The lowest curve in FIG. 2 shows the ground state energy divided by $n$ as a function of $t$.
(Since the system is decoupled this is actually the ground state energy of a single qubit.) 
We then consider the $n$ bit scrambled problem for different values of $n$. At each $n$ we pick
a single random permutation $\pi$ of $0,\dots,(2^n-1)$ and apply it to obtain a cost 
function $h(\pi^{-1}(z))$ while keeping $H_B$ fixed. 
The ground state energy divided by $n$ is now plotted for $n=9,12,15$ and $18$.
From these scrambled problems it is clear that if we let $n$ get large the typical curves 
will approach a triangle with a discontinuous first derivative at $t=T/2$. For large $n$, the
ground state changes dramatically as $t$ passes through $T/2$. In order to keep the quantum system
in the ground state we need to go very slowly near $t=T/2$ and this results in a long required run time.


\section{Conclusions}

In this paper we have two main results about the performance of the quantum adiabatic algorithm when used
to find the minimum of a classical cost function $h(z)$ with $z=0,\dots,N-1$. Theorem 1 says that 
for any cost function $h(z)$, if the beginning Hamiltonian is a one dimensional projector onto the
uniform superposition of all the $\ket{z}$ basis states, the algorithm will not find the minimum
of $h$ if $T$ is less then of order $\sqrt{N}$. This is true regardless of how simple it is to
classically find the minimum of $h(z)$.

In Theorem 2 we start with any beginning Hamiltonian and classical cost function $h$. Replacing $h(z)$
by a scrambled version, i.e. $h^{[\pi]}(z)=h(\pi(z))$ with $\pi$ a permutation of $0$ to $N-1$,
will make it impossible for the algorithm to find the minimum of $h^{[\pi]}$ in time less than
order $\sqrt{N}$ for a typical permutation $\pi$. 
For example suppose we have a cost function $h(z)$ and have chosen $H_B$ so that
the quantum algorithm finds the minimum in time of order $\textrm{log}\,N$. Still scrambling
the cost function results in algorithmic failure.

These results do not imply anything about the more interesting case where $H_B$
and $H_P$ are structured, i.e., sums of terms each operating only on several qubits.


\section*{Acknowledgements}
The authors gratefully acknowledge support from the National Security Agency (NSA) and Advanced Research 
and Development Activity (ARDA) under Army Research Office (ARO) contract W911NF-04-1-0216.


\appendix

\section{Transitions in a two level system}

Let us consider a two level system with Hamiltonian
\begin{eqnarray*}
  H(s) = E_0(s) \ket{\phi_0(s)}\bra{\phi_0(s)} + E_1(s) \ket{\phi_1(s)}\bra{\phi_1(s)},
\end{eqnarray*}
which varies smoothly with $s=t/T$. Here $\ket{\phi_0(s)}$ and $\ket{\phi_1(s)}$ are orthonormal for all $s$.
The Schr\"{o}dinger equation reads
\begin{eqnarray*}
  i \deriv{}{s} \ket{\psi} = T H(s) \ket{\psi}.
\end{eqnarray*}
The two energy levels in the system are separated by a gap 
\[
	g(s)=E_1(s)-E_0(s),
\]
which we assume is always larger than $0$. Let us introduce $\theta$ (with the dimension of energy) as 
\[
	\theta(s) = \int_0^s g(s') \,\textrm{d}s',
\]
and let
\begin{eqnarray}
	\ket{\psi(s)} = c_0(s) e^{-iT \int_0^s E_0(s')\,\textrm{d}s'}\ket{\phi_0(s)} + 
		c_1(s) e^{-iT \int_0^s E_1(s')\,\textrm{d}s'}\ket{\phi_1(s)}. 		\label{statesolution}
\end{eqnarray}
We pick the phases of $\ket{\phi_1(s)}$ and $\ket{\phi_0(s)}$ such that 
$\bra{\phi_1(s)}\deriv{}{s}\ket{\phi_1(s)}=\bra{\phi_0(s)}\deriv{}{s}\ket{\phi_0(s)}=0$.
Plugging (A1) into the Schr\"odinger equation gives
\begin{eqnarray*}
	\deriv{c_0}{s} &=& c_1 e^{-iT\theta} \Big\langle \phi_1 \Big| \deriv{}{s} \Big| \phi_0\Big\rangle^*, \\
	\deriv{c_1}{s} &=& -c_0 e^{iT\theta} \Big\langle \phi_1 \Big| \deriv{}{s} \Big| \phi_0\Big\rangle,
\end{eqnarray*}
or equivalently,
\begin{eqnarray*}
	\deriv{c_0}{\theta} &=& c_1 e^{-iT\theta} f^*, \\
	\deriv{c_1}{\theta} &=& -c_0 e^{iT\theta} f,
\end{eqnarray*}
where 
\[
	f(\theta) \equiv \Big\langle \phi_1 \Big| \deriv{}{\theta} \Big| \phi_0\Big\rangle
		 = -\frac{1}{g} \Big\langle \phi_1 \Big| \deriv{H}{\theta} \Big| \phi_0\Big\rangle
		 = -\frac{1}{g^2} \Big\langle \phi_1 \Big| \deriv{H}{s} \Big| \phi_0\Big\rangle.
\]
Now let $\theta(1)=\bar{\theta}$. 
We have $c_1(0)=0$ and we want the transition amplitude at $s=1$ which is 
\begin{eqnarray*}
   c_1(\bar{\theta}) &=& -\int_0^{\bar{\theta}} c_0 e^{iT\theta}f\,\textrm{d}\theta \\
 	&=& \left[-c_0 f \frac{e^{iT\theta}}{iT}\right]^{\bar{\theta}}_0 + 
   \frac{1}{iT} \int_0^{\bar{\theta}} e^{iT\theta} \left( \deriv{c_0}{\theta}f+ c_0 \deriv{f}{\theta}\right) \textrm{d}\theta \\
 	&=& \frac{1}{T}\left( \left[i c_0 f e^{iT\theta} \right]^{\bar{\theta}}_0 
    -i \int_0^{\bar{\theta}} \left( c_1 f f^*+ e^{iT\theta} c_0 \deriv{f}{\theta} \right) \textrm{d}\theta \right). 
\end{eqnarray*}
Now $|c_0|\leq 1$ and $|c_1|\leq 1$. 
As long as the gap does not vanish $|f(\theta)|$ and $\left|\deriv{f}{\theta}\right|$ are bounded so we have that
$\left|c_1(\bar{\theta})\right| = O\left(\frac{1}{T}\right)$.
The probability of transition to the excited state for a two-level system with a nonzero gap is thus 
\[
     \left|c_1(\bar{\theta})\right|^2 = O(T^{-2}).
\]


\end{document}